\documentclass[conference]{IEEEtran}

\usepackage{color}

\usepackage{amsmath}%
\usepackage{amsfonts}%
\usepackage{amssymb}%
\usepackage{graphicx}
\usepackage{subfig}

\newtheorem{theorem}{Theorem}

\newtheorem{lemma}[theorem]{Lemma}

\newenvironment{proof}[1][Proof]{\textbf{#1.} }{\ \rule{0.5em}{0.5em}}

\begin{document}
\title{The MIMO Wireless Switch: \\ Relaying Can Increase the Multiplexing Gain}
\author{
\authorblockN{Hassan Ghozlan, Yahya Mohasseb}
\authorblockA{Wireless Intelligent Networks Center (WINC)\\
Nile University, Cairo, Egypt\\
hassan.ghozlan@nileu.edu.eg\\
yahya.mohasseb@nileuniversity.edu.eg}
\and
\authorblockN{Hesham El Gamal}
\authorblockA{Department of Electrical\\
and Computer Engineering\\
Ohio State University, Columbus, OH, USA\\
helgamal@ece.osu.edu}
\and
\authorblockN{Gerhard Kramer}
\authorblockA{Bell Labs\\
Alcatel-Lucent\\Murray Hill, NJ, USA\\
gkr@research.bell-labs.com}
}
\maketitle

\begin{abstract}
This paper considers an interference network composed of $K$ half-duplex single-antenna pairs of users who wish to establish bi-directional communication with the aid of a multi-input-multi-output (MIMO) half-duplex relay node. This channel is referred to as the ``MIMO Wireless Switch'' since, for the sake of simplicity, our model assumes no direct link between the two end nodes of each pair implying that all communication must go through the relay node (i.e., the MIMO switch). Assuming a delay-limited scenario, the fundamental limits in the high signal-to-noise ratio (SNR) regime is analyzed using the diversity-multiplexing tradeoff (DMT) framework. Our results sheds light on the structure of optimal transmission schemes and the gain offered by the relay node in two distinct cases, namely reciprocal and non-reciprocal channels (between the relay and end-users). In particular, the existence of a relay node, equipped with a sufficient number of antennas, is shown to increase the multiplexing gain; as compared with the traditional fully connected $K$-pair interference channel. To the best of our knowledge, this is the first known example where adding a relay node results in enlarging the pre-log factor of the sum rate. Moreover, for the case of reciprocal channels, it is shown that, when the relay has a number of antennas at least equal to the sum of antennas of all the users, static time allocation of decode and forward (DF) type schemes is optimal. On the other hand, in the non-reciprocal scenario, we establish the optimality of dynamic decode and forward in certain relevant scenarios.
\end{abstract}

\section{Introduction}

It is natural to expect next-generation communication networks to require higher data rates and offer more guarantees on the Quality of Service (QoS) metrics, in terms of both delay and robustness to noise and interference. User cooperation and relaying is one of the most promising techniques for meeting these new challenges. 
The recent work of \cite{Sendonaris2003a} and \cite{Sendonaris2003} triggered a vast literature on developing cooperative relaying techniques, e.g. \cite{Yuksel2007} and references therein. 
Here, we consider an extended {\em multi-user} version of the two-way relay channel. More specifically, a $K$-pair interference network is analyzed. Each pair wishes to establish a two-way communication link in the presence of interference from the other pairs. Only one multi-input-multi-output (MIMO) relay is responsible for helping all the $K$ pairs. To simplify the analysis, we further limit our analysis to the multi-hop scenario where all the communication must go through the relay node. Quite remarkably, our results establish the ability of the relay node, i.e., switch in our multi-hop set-up, to increase the pre-log factor of the sum rate as compared with that of the $K$ pair interference channel (with direct links) in certain special cases.

Our analysis focuses on the delay-limited high SNR regime. Towards this end, we characterize the diversity-multiplexing tradeoff~\cite{Zheng2003} of the MIMO switch channel under two different assumptions on the channel reciprocity between the relay node and end-users. Our results shed light on the structure of the optimal schemes, the gain offered by the relay node, and the critical impact of channel state information (CSI) on the problem. 

The rest of the paper is organized as follows. The system model and our notations are described in Section \ref{sec:SystemModel} while the main results are presented in Section \ref{sec:MainResults}. Finally, Section \ref{sec:Discussion} concludes the paper with a brief discussion outlining the main insights gleaned from our results.

%++++++++++++++++++++++++++++++++++++++++++++++++++++++++++++++++++++++++++++++++++++++++++++++++++++++++++++++++++++++++++++++++++++++++++++++++++++++
\section{System Model and Notation}
\label{sec:SystemModel}

Throughout the paper, we write $f(\rho) \doteq \rho^b$ if
$\lim_{\rho \rightarrow \infty} \frac{\log f(\rho)}{\log(\rho)} = b$
where $b$ is called the exponential order of $f(\rho)$.
The inequalities $\stackrel{\cdot}{\geq}$ and $\stackrel{\cdot}{\leq}$ are defined similarly.
We use $(x)^+$to mean $\max\{x,0\}$.
$\mathbb{R}^N$ and $\mathbb{C}^N$ denote the set of real and complex $N$-tuples respectively.
We denote the complement of set $\mathcal{O}$ by $\overline{\mathcal{O}}$ whereas
the cardinality of a set $\Lambda$ is denoted by $|\Lambda|$.
$\log(\cdot)$ denotes the base-$2$ logarithm,
$\otimes$ denotes Kronecker's multiplication, 
and $^\dagger$ denotes Hermitian transpose.

We consider a network composed of $K$ pairs of users and a relay $R$. Each user is equipped with a single antenna while the relay is equipped with $M$ antennas. In our model, there is no direct link between the users. The received signal at the relay over $T_1$ channel uses is given by
\begin{equation}
	\boldsymbol{Y}_{r} =
	\sqrt{\rho} \sum_{k=1}^{K} \sum_{i=1}^{2} \left(\boldsymbol{I}_{T_1} \otimes \boldsymbol{H}_{k,i}^{(1)}\right) \boldsymbol{X}_{k,i} + \boldsymbol{W}_r^{(1)},
\end{equation}
whereas the received signal by user $U_{k,i}$ over $T_2$ channel uses is given by
\begin{equation}
	\boldsymbol{Y}_{k,i} =
	\sqrt{\frac{\rho}{M}} \left(\boldsymbol{I}_{T_2} \otimes \boldsymbol{H}_{k,i}^{(2)}\right) \boldsymbol{X}_{k,r} + \boldsymbol{W}_{k,i}^{(2)}.
\end{equation}

In this notation, 
$\rho$ is the average SNR per link,
$\boldsymbol{Y}_{r} \in \mathbb{C}^{M T_1 \times 1}$ is the received signal at the relay,
$\boldsymbol{X}_{k,i} \in \mathbb{C}^{T_1 \times 1}$  is the node $U_{k,i}$ input matrix,
$\boldsymbol{Y}_{k,i} \in \mathbb{C}^{T_2 \times 1}$ is the received signals by node $U_{k,i}$ from the relay,
and $\boldsymbol{X}_r \in \mathbb{C}^{M T_2 \times 1}$ is the relay input matrix. 
All the channels are assumed to be frequency non-selective, quasi-static Rayleigh fading and independent of each other; that is $\boldsymbol{H}_{k,i}^{(j)}$, $j=1,2$, is a matrix whose entries are independent and identically distributed (i.i.d) complex circularly symmetric Gaussian random variables with zero mean and unit variance i.e., $\mathcal{N}_\mathbb{C}(0,1)$.
The additive white Guassian noise $\boldsymbol{W}_r^{(1)}$ and $\boldsymbol{W}_{k,i}^{(2)}$ have i.i.d entries with $\mathcal{N}_\mathbb{C}(0,1)$
where $\boldsymbol{W}_r^{(1)} \in \mathbb{C}^{M T_1\times 1}$ and $\boldsymbol{W}_{k,i}^{(2)} \in \mathbb{C}^{T_2 \times 1}$.

All the nodes are assumed to operate in half-duplex mode, i.e., at any point in time, a node can only listen or transmit but not both.
We consider a short-term (or per-block) average power constraint:
$\mbox{tr}\left( E[ \boldsymbol{X}_{k,i}^\dagger \boldsymbol{X}_{k,i} ]\right) \leq T_1$ and $\mbox{tr}\left( E[ \boldsymbol{X}_{k,r}^\dagger \boldsymbol{X}_{k,r} ]\right) \leq M T_2$. 
We focus on the {\bf symmetric case} with equal rates assigned to all the $2K$ sources in the network. For increasing SNR $\rho$, we say that a scheme achieves \textit{ multiplexing gain} $r$ {\bf per each user} in the network and \textit{diversity gain} $d$ if the rate $R$ (per user) and the average error probability $P_E$ satisfy
\begin{equation}
	\lim_{\rho \rightarrow \infty} \frac{R(\rho)}{\textrm{log}(\rho)} = r
	\mbox{ and }
	\lim_{\rho \rightarrow \infty} \frac{\textrm{log} P_E(\rho)}{\textrm{log}(\rho)} = -d
\end{equation}
Therefore, the multiplexing gain {\bf per pair} is $2r$ and {\bf sum multiplexing gain} is $2Kr$.  For the sake of completeness, we state the results obtained in \cite{Tse2004} regarding the diversity-multiplexing tradeoff of the multiple access channel (MAC).
If the block length $l \geq Km + n -1$, the optimal tradeoff curve for a MAC of $K$ users (with symmetric rates of $r \log\rho$) each with $m$ antennas and a receiver with $n$ antennas is given by
\begin{equation}
	d^{MAC-sym}_{K,m,n}(r) = \left\{
	\begin{array}{ll}
	d^{PPC}_{m,n}(r) ,   &   r \leq \min(m,\frac{n}{K+1})  \\
	d^{PPC}_{Km,n}(Kr) , &   r \geq \min(m,\frac{n}{K+1})
	\end{array}
	\right.
\end{equation}
where $d^{PPC}_{m,n}(r)$ is the optimal tradeoff curve $d^{PPC}_{m,n}(r)$ for a point-to-point channel with $m$ transmit antennas and $n$ receive antennas~\cite{Zheng2003}.

Finally, we consider two distinct cases for the channels between the relay and the users. In the reciprocal channels case, the channel from the user to the relay is the reciprocal of the channel from the relay to the user i.e $\boldsymbol{H}_{k,i}^{(1)} = \boldsymbol{H}_{k,i}^{(2)} = \boldsymbol{H}_{k,i} $ for all $i$ and $k$. In the non-reciprocal independent channels case, the channels between the users and the relay in both direction are independent of each other i.e $\boldsymbol{H}_{k,i}^{(1)}$ and $\boldsymbol{H}_{k,i}^{(2)}$ are independent.
Practically, one would expect the channels to be correlated. However, we only consider the two extreme cases to allow for obtaining insights without complicating the analysis. Perfect knowledge of Channel State Information (CSI) is only assumed at the receivers. However, in the reciprocal channel scenario, the receiver CSI also implies transmitter CSI at the relay node. As shown next, this knowledge can be exploited to obtain significant performance gains.

\section{Main Results}
\label{sec:MainResults}

In our achievability arguments, we use the following two schemes: 1)DF-MAC-TDMA and 2)DF-MAC-BC. Both schemes operate in two phases because of the half-duplex constraint. For the DF-MAC-TDMA scheme, \textbf{Phase One} is a multiple-access phase in which each of the sources sends its message to the relay using codebooks of rate $R$. Our relay (switch) decodes the messages jointly then XORs the messages of each pair and encodes the new messages with an independent codebook of rate $R$.
In \textbf{Phase Two}, the relay transmits on $K$ TDMA slots, where the $k$-th slot is allocated for the the message intended for the $k$-th pair of users, $U_{k,1}$ and $U_{k,2}$. The DF-MAC-BC scheme differs from the DF-MAC-TDMA scheme only in \textbf{Phase Two} where the relay transmits to all the pairs simultaneously using the available transmit CSI in the reciprocal channel scenario. In the {\bf static} version of our two schemes, the time interval allocated to each phase is fixed {\em a-priori} whereas the {\bf dynamic} version allows for changing the allocation based on the instantaneous realizations of the channel matrices.

\subsection{Reciprocal Channels}
\begin{theorem}
The diversity-multiplexing tradeoff of the $K$-Pair MIMO Switch channel with reciprocal channels is bounded by
\begin{align}
	d^{MAC-sym}_{2K,1,M}\left(\frac{r}{a^*}\right) \leq d(r) \leq M\left(1-2 r\right)^+ 
\end{align}
where $a^*$ satisfies
\begin{equation}
	d^{MAC-sym}_{2K,1,M}(r) \left(\frac{r}{a^*}\right) = d^{MAC-sym}_{K,1,M} \left(\frac{r}{1-a^*} \right) 
	\label{eq:sdf_macbc_condition}
\end{equation}

When $M \geq 2K$, the lower bound matches the upper bound yielding the optimal DMT. Moreover, the static DF-MAC-BC scheme achieves the optimal DMT in this special case.
\end{theorem}

\begin{proof} (Sketch)
We start with the achievability of the static DF-MAC-BC scheme for $K$ pairs.
We define the error event as the error in decoding a message of one user (in any pair) at the other user (in the same pair), i.e.,
\begin{equation}
	E = \bigcup_{k=1,\cdots,K} \bigcup_{i=1,2} \left\{\hat{m}_{k,i} \neq m_{k,i} \right\}
\end{equation}
where
$m_{k,i}$ is the message of $U_{k,i}$.
$\hat{m}_{k,1}$ is the estimate of $m_{k,1}$ at $U_{k,2}$ and
$\hat{m}_{k,2}$ is the estimate of $m_{k,2}$ at $U_{k,1}$.
Using Bayes' rules, we can upper bound $P_E$ as
\begin{equation}
	P(E) \leq  P(E|\overline{E}_r)  + P(E_r)
\end{equation}
where $E_r$ is the error event at the relay.

The probability of error in Phase One, $P(E_r)$, is that of a multiple-access channel of $2K$ single-antenna users having symmetric rates of $\frac{r}{a}$ and a receiver having $n$ antennas. Hence, the DMT is  given by $d^{MAC-sym}_{2K,1,M} \left(\frac{r}{a} \right)$.

The probability of error in Phase Two, $P(E|\overline{E}_r)$, has an exponential order $-d^{BC-sym}_{K,1,M} \left(\frac{r}{1-a} \right)$ 
where $d^{BC-sym}_{K,m,n}(r)$
is the optimal tradeoff of a broadcast channel with a transmitter, having $n$ transmit antennas, transmitting
to $K$ users (each with $m$ receive antennas)  with individual rates $r \log\rho$ .
%\comment{or...broadcast channel with a transmitter, having $n$ transmit antennas, transmitting $K$ independent messages with rate $(K?)r \log\rho$ to users with $m$ receive antennas.}
Therefore, $d_{MAC-BC}(r)$ is lower bounded by
\begin{align}
	\max_{a} \min \bigg\{ d^{MAC-sym}_{2K,1,M}(r) \left(\frac{r}{a} \right)  , d^{BC-sym}_{K,1,M} \left(\frac{r}{1-a} \right) \bigg\}
\end{align}

The duality between the multiple-access channel and broadcast channels in \cite{JindalMay2004} implies that they have the same optimal tradeoff curve i.e
$d^{MAC-sym}_{K,m,n}(r) = d^{BC-sym}_{K,m,n}(r)$. The optimal value for static time allocation, $a^*$, is obtained when the diversity gains of both phases are equal,  yielding equation (\ref{eq:sdf_macbc_condition}). Hence,
\begin{equation}
	d_{MAC-BC}(r) \geq  d^{MAC-sym}_{2K,1,M} \left(\frac{r}{a^*} \right)
\end{equation}

%---------------------------------------------------------------------------------------------------------------------------------------------
This completes the achievability. We now move to the converse.
We use a genie-aided strategy to lower bound the probability of error.
In phase one, all the messages except one message, say $m_{1,1}$, are revealed to the relay, 
while in phase two, all the messages except the message intended for user $U_{1,1}$ are revealed to their destinations.

Thus we obtain two lower bounds on the probability of error
$P_{E}(\rho|H) \geq P_{\tilde{m}_{1,1} \neq m_{1,1}}(\rho|H) $
and
$P_{E}(\rho|H) \geq P_{\hat{m}_{1,2} \neq \tilde{m}_{1,2}}(\rho|H) $, 
where $\tilde{m}_{k,i}$ is the estimate of $m_{k,i}$ at $R$.
	
This means that $P_{E}(\rho|H)$ is lower bounded by the maximum of the two bounds. We minimize this maximum to tighten the lower bound and use Fano's inequality to get
\begin{equation}
	P_{E}(\rho|H) \geq
	1 - \frac{1}{r T \log\rho} \mathcal{I} - \frac{1}{r T \log\rho}
\end{equation}
where
\begin{equation}
	\mathcal{I} = \max_{a,P_{X_{1,1}},P_{X_{r}}} \min \left\{I_H(X_{1,1};Y_r),I_H(X_{r};Y_{1,1})\right\}
	\label{eq:max_min_mutual_info}
\end{equation}
in which $P_{X_{1,1}}$ and $P_{X_{r}}$ are the probability distribution functions of $X_{1,1}$ and $X_r$ respectively.
Using Gaussian inputs to maximize the mutual information leads to
\begin{align}
	\mathcal{I} &= \max_{a} \min \left\{a T C_1 , (1-a)T C_1\right\}	
\end{align}
where $C_1 = \log \det (I + \rho H_{1,1} H_{1,1}^{\dagger})$.
Clearly, the optimal value for $a$ is $0.5$. Averaging over all the channel realizations yields
\begin{align}
	P_{E}(\rho) \stackrel{\cdot}{\geq}	P(\mathcal{I}< R T) = P\left( C_1 /2< R \right)
\end{align}

$P\left( C_1 < 2R \right)$ has an exponential order of $-d^{PPC}_{M,1}(2r)$. Hence,
$P_E \stackrel{\cdot}{\geq } {\rho}^{-M(1-2 r)^+}$ and consequently,
$d(r)$ is upper bounded by $d(r) {\leq } M(1-2 r)^+$
\end{proof}

In the previous result, the channel reciprocity played a key role in offering the relay node transmitter CSI. In some practical scenarios, it maybe desirable to use {\bf robust} protocols that do not depend on the availability of such side information. The following Lemma characterizes the performance of one of such schemes. Even in this scenario, this result establishes the ability of the relay node to increase the maximum multiplexing gain as compared with the $K$-pair interference channel. Interestingly, using this scheme one obtains the maximum multiplexing gain {\bf per pair} in the special case $K=2$. 

\begin{lemma}
The diversity-multiplexing tradeoff of the static DF-MAC-TDMA scheme for $K$ pairs of users is lower bounded by
\begin{align}
d^{MAC-sym}_{2K,1,M} \left(\frac{r}{a^*}\right) \leq d_{MAC-TDMA}(r)
\end{align}
where $a^*$ satisfies
\begin{equation}
d^{MAC-sym}_{2K,1,M} \left(\frac{r}{a^*} \right)  = d^{PPC}_{M,1} \left(\frac{Kr}{1-a^*} \right)
\label{eq:sdf_mactdma_condition}
\end{equation}

\end{lemma}

\begin{proof} (Sketch)
The achievability of the static DF-MAC-TDMA scheme for $K$ pairs follows the same steps of static DF-MAC-BC scheme.
However, the probability of error in Phase Two, $P(E|\overline{E}_r)$, is dominated by the worst (i.e maximum) probability of error of the point-to-point links between the relay and the users. Because of symmetry, these probabilities have the same exponential order of decay with SNR which is $d^{PPC}_{M,1} \left(\frac{Kr}{1-a} \right)$. 
Consequently, the lower bound on $d_{MAC-TDMA}(r)$ is given by
\begin{equation}
	\max_{a} \min \left\{ d^{MAC-sym}_{2K,1,M} \left(\frac{r}{a} \right)  , d^{PPC}_{M,1} \left(\frac{Kr}{1-a} \right) \right\}
\end{equation}
The optimal value, $a^*$, is obtained when the diversity gains of both phases are equal, giving equation (\ref{eq:sdf_mactdma_condition}).
%\begin{equation}
%	d^{MAC-sym}_{2K,1,M} \left(\frac{r}{a^*} \right)  = d^{PPC}_{M,1} \left(\frac{Kr}{1-a^*} \right)
%\end{equation}
\end{proof}

%-----------------------------------------------------------------------------------------------------------------------------------------------------
%-----------------------------------------------------------------------------------------------------------------------------------------------------
\subsection{Non-Reciprocal Independent Channels}
Here, only receiver CSI is available at the relay. Moreover, the lack of channel reciprocity makes the dynamic version of our protocol superior to the static version. The following result formalizes this observation and provides upper and lower bounds on the optimal diversity-multiplexing tradeoff.

\begin{theorem}
The diversity-multiplexing tradeoff of the $K$-Pair MIMO Switch channel with independent channels is bounded by
\begin{align}
	d_{DDF}(r) \leq d(r) \leq M\left(\frac{1-(K+1)r}{1-r}\right)^+
	\label{eq:ddf_bounds}
\end{align}
%-----------------------------------------------------------------------------------------------------------------------------------------
where
\begin{equation}
	d_{DDF}(r) =  \min_{\Lambda} \inf_{(\alpha_1,\alpha_2) \in \tilde{\mathcal{O}}_2^\Lambda}
	\sum_{i=1}^{2}
	\sum_{j=1}^{M_i^*}
	(2j-1+|M_i-M_{i+1}|) \alpha_{i,j}
	\label{eq:d_ddf}
\end{equation}
for $\Lambda \subseteq \{(k,i)|k=1,\cdots,K,i=1,2\}$ and
\begin{align*}
\tilde{\mathcal{O}}_2^{\Lambda} = \bigg\{ &
(\alpha_1,\alpha_2) \in \mathbb{R}^{M_1^*+} \times \mathbb{R}^{M_2^*+} |
\alpha_{i,1} \geq \ldots \geq \alpha_{i,M_i^*} \geq 0, \\ &
\frac{S_1^{\Lambda} S_2}{K S_1^{\Lambda} + |\Lambda| S_2}  < r
\bigg\}		
\end{align*}
in which we have $M_1 = |\Lambda|$, $M_2 = M$, $M_3 = 1$, $M_i^* = \min\{M_i,M_{i+1}\}$.
Whereas $S_1^{\Lambda} \stackrel{\Delta}{=} \sum_{j=1}^{\min\{|\Lambda|,M\}}  \left(1-\alpha^{\Lambda}_{1,j}\right)^+$
and $S_2 \stackrel{\Delta}{=} \left(1-\alpha_{2,1}\right)^+$.
%-----------------------------------------------------------------------------------------------------------------------------------------

When $M \geq 2K$, the lower bound matches the upper bound yielding the optimal DMT. Moreover, dynamic DF-MAC-TDMA scheme achieves this DMT.

\end{theorem}

\begin{proof} (Sketch)
%#####################################################################################################################################################
First, we consider the achievability of the dynamic DF scheme.
The relay listens for $aT$ channel uses until the messages of all the users can be decoded
\begin{equation}
	\frac{1}{T} I_H(X_{\Lambda};Y_r|X_{\overline{\Lambda}}) > |\Lambda| R
\end{equation}
where $\Lambda \subseteq \{(k,i)|k=1,\cdots,K,i=1,2\}$.
Using Gaussian inputs, the previous equation reduces to $a C_1^{\Lambda} > |\Lambda| R$
where $C_1^{\Lambda} = \log \det \left(I + H_{\Lambda}^{(1)} H_{\Lambda}^{(1) \dagger} \right)$.
$H_{\Lambda}^{(1)}$ is a matrix augmenting the channel matrices from the users in the set $\Lambda$ to the relay.
Therefore, we have
$a = \max_{\Lambda} \left\{ \frac{|\Lambda| R}{C_1^{\Lambda}} \right\}$.
If $a>1$ then the whole system is in outage. We define this event as $\mathcal{O}_1 \stackrel{\Delta}{=} \{a>1\}$ and its probability, $P(\mathcal{O}_1)$, is equal to
\begin{align}
	P\left( \bigcup_{\Lambda} \left\{ \frac{|\Lambda| R}{C_1^{\Lambda}} > 1 \right\} \right)
	\doteq \max_{\Lambda} P\left( \frac{S_1^{\Lambda}}{|\Lambda|} < r \right)
\end{align}
where $S_1^{\Lambda} \stackrel{\Delta}{=} \sum_{j=1}^{\min\{|\Lambda|,M\}}  \left(1-\alpha^{\Lambda}_{1,j}\right)^+$.

If the decoding was successful at the relay, i.e., $a<1$
then, the outage occurs if the mutual information between the transmitted signal $X_r$ and the received signal $Y_{k,i}$ of user $U_{k,i}$ does not support the target data rate. We will focus only on user $U_{1,1}$ since all users have the same outage behavior. Therefore, outage occurs when
$\frac{1}{T} I_H(X_r;Y_{1,1}) < R$.
Again, we assume $X_r$ to be Guassian and we define this outage event as $\mathcal{O}_2 \stackrel{\Delta}{=} \{(1-a)C_2/K < R\}$ where $C_2 = \log \det \left(I + H_{1,1}^{(2)} H_{1,1}^{(2) \dagger} \right)$.
\begin{align}
	P \left( \frac{1-a}{K} C_2 < R  \right)
	& = P\left(  \bigcup_{\Lambda} \left\{ \frac{C_1^{\Lambda} C_2}{K C_1^{\Lambda} + |\Lambda| C_2}   < R \right\} \right) \nonumber \\
	& \doteq \max_{\Lambda} P\left(   \frac{S_1^{\Lambda} S_2}{K S_1^{\Lambda} + |\Lambda| S_2}   < r \right)
\end{align}
where $S_2 \stackrel{\Delta}{=} \sum_{j=1}^{\min\{M,1\}}  \left(1-\alpha_{2,j}\right)^+$.

The overall probability of outage is given by
\begin{align}
	P(\mathcal{O}_1 \cup \mathcal{O}_2)
	& = P(a>1) + P \left( \frac{1-a}{K} C_2 < R , a<1 \right) \nonumber \\
	& \doteq \max_{\Lambda} P\left( \frac{S_1^{\Lambda} S_2}{K S_1^{\Lambda} + |\Lambda| S_2} < r \right) \nonumber \\
	& \doteq \max_{\Lambda} P\left(   \mathcal{O}^{\Lambda}_2 \right) \doteq \max_{\Lambda} \rho^{-d^{\Lambda}(r)}	\doteq \rho^{-d_{DDF}(r)}
\end{align}
where
\begin{align*}
\mathcal{O}_2^{\Lambda} = \bigg\{ &
(\alpha_1,\alpha_2) \in \mathbb{R}^{M_1^*} \times \mathbb{R}^{M_2^*} |
\alpha_{i,1} \geq \ldots \geq \alpha_{i,M_i^*} \geq 0, \\ &
\frac{S_1^{\Lambda} S_2}{K S_1^{\Lambda} + |\Lambda| S_2}  < r
\bigg\}		
\end{align*}
and
%-----------------------------------------------------------------------------------------------------------------------------------------
\begin{equation}
	d^{\Lambda}(r) = \inf_{(\alpha_1,\alpha_2) \in \tilde{\mathcal{O}}_2^\Lambda}
	\sum_{i=1}^{2}
	\sum_{j=1}^{M_i^*}
	(2j-1+|M_i-M_{i+1}|) \alpha_{i,j}
\end{equation}
in which we have $M_1 = |\Lambda|$, $M_2 = M$, $M_3 = 1$, $M_i^* = \min\{M_i,M_{i+1}\}$
and $ \tilde{\mathcal{O}}_2^{\Lambda} = {\mathcal{O}}_2^{\Lambda} \cap \mathbb{R}^{M_1^*+} \times \mathbb{R}^{M_2^*+}$.

%-----------------------------------------------------------------------------------------------------------------------------------------
Finally, 
\begin{equation}
	d_{DDF}(r) = \min_{\Lambda} d^{\Lambda}(r)
\end{equation}
%#####################################################################################################################################################

The converse, in this case, follows the same outline of the converse in the case of reciprocal channels. 
However, using Gaussian inputs gives a different expression for $\mathcal{I}$, namely,
\begin{equation}
	\mathcal{I} = \max_{a} \min \left\{aT C_1^{(1)},\frac{(1-a)T}{K}C_1^{(2)}\right\}
\end{equation}
where $C_1^{(i)} = \log \det (I + \rho H_{1,1}^{(i)} {H_{1,1}^{(i)}}{^\dagger})$.
The optimal choice of $a$ is
$\frac{C_1^{(2)}}{K C_1^{(1)} + C_1^{(2)}}$.
By averaging over all channel realizations, we obtain
\begin{equation}
	P_{E}(\rho) \stackrel{\cdot}{\geq}	P(\mathcal{I}< R T)
	= P \left( \frac{C_1^{(1)} C_1^{(2)}}{K C_1^{(1)} + C_1^{(2)}} < R\right)
\end{equation}

At high SNR, $C_1^{(i)}$ can be characterized as
$C_1^{(i)} \doteq \log\rho^{S_1^{(i)}}$
where $S_1^{(i)} = \left( 1-\alpha_1^{(i)} \right)^+$ and $i=1,2$.

\begin{equation}
 P(\mathcal{I}< R T) \doteq P \left( \frac{S_1^{(1)} S_1^{(2)}}{K S_1^{(1)} + S_1^{(2)}} < r\right) \doteq \rho^{-d_{out}(r)}
\end{equation}

\begin{equation}
	d_{out}(r) = \inf_{\left(\alpha_1^{(1)},\alpha_1^{(2)}\right) \in \tilde{\mathcal{O}}}
	\sum_{i=1}^{2}
	M \alpha_1^{(i)}
\end{equation}

\begin{align*}
\tilde{\mathcal{O}} = \left\{
\left( \alpha_1^{(1)},\alpha_1^{(2)} \right) \in \mathbb{R}^{+} \times \mathbb{R}^{+} \left|
\frac{S_1^{(1)} S_1^{(2)}}{K S_1^{(1)} + S_1^{(2)}}  < r
\right. \right\}		
\end{align*}

Solving the optimization problem at hand yields the upper bound in (\ref{eq:ddf_bounds}).
\end{proof}

\section{Discussion}
\label{sec:Discussion}
\begin{figure}
	\centering
		\includegraphics[width=0.45\textwidth]{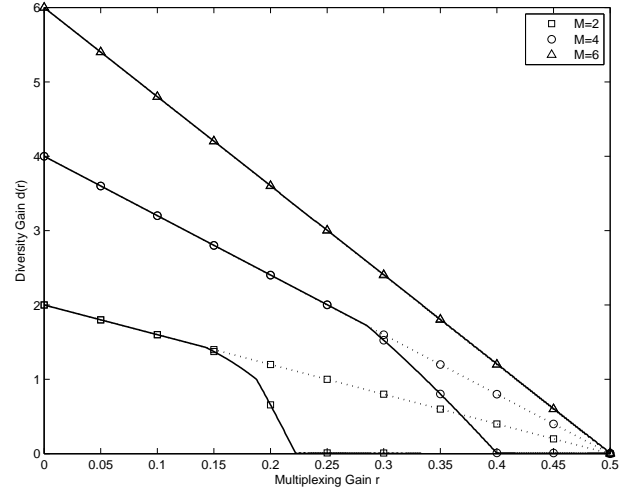}
	\caption{Lower and upper bounds on the diversity-multiplexing tradeoff for static DF-MAC-BC ($K=3$).}
\label{fig:sdf3pairs}
\end{figure}

\begin{figure}
	\centering
		\includegraphics[width=0.45\textwidth]{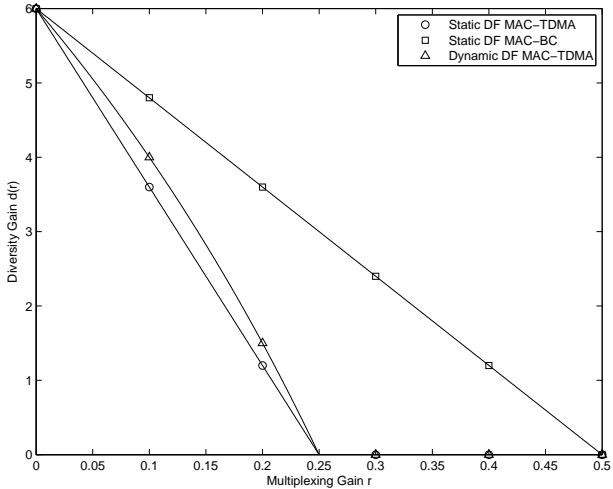}
	\caption{The diversity-multiplexing tradeoff of different schemes in the case of $3$ pairs and a relay with $6$ antennas.}
	\label{fig:schemes3pairs}
\end{figure}
\begin{figure}
	\centering
		\includegraphics[width=0.45\textwidth]{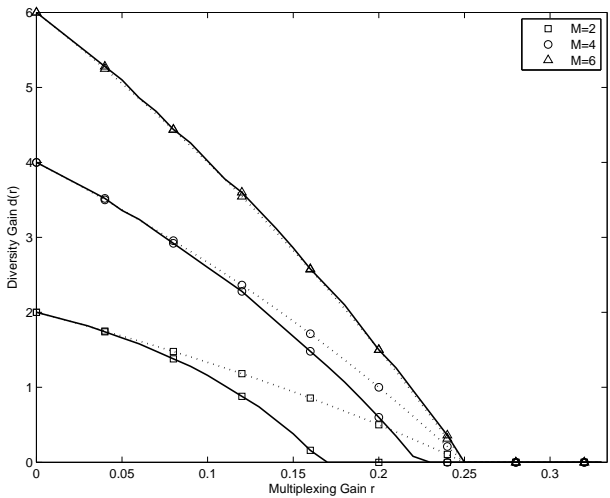}
	\caption{Lower and upper bounds on the diversity-multiplexing tradeoff for the dynamic DF-MAC-TDMA ($K=3$).}
	\label{fig:ddf3pairs}
\end{figure}

\begin{enumerate}
\item  In the case of the reciprocal channels, Figure \ref{fig:sdf3pairs} shows that the lower bound (solid lines) matches the upper bound (dashed lines) when the number of relay antennas is equal to total number of antennas of users i.e $M=2K$. For $M<2K$, there is a gap between the two bounds for high multiplexing gains. Similar behavior is shown in Figure \ref{fig:ddf3pairs} for the case of identical channels.

\item DF-MAC-TDMA is shown to be suboptimal in the case of reciprocal channels whereas dynamic DF-MAC-TDMA is optimal (for $M\geq2K$) in the case of non-reciprocal channels. This can be explained by the fact that when Transmit CSI is not available then, TDMA is optimal \cite{Caire2003} which is indeed the case for the identical channels scenario. However, in the case of reciprocal channels, the receive CSI assumption, coupled with reciprocity, implies that Transmit CSI is available at the relay. Hence, {\bf simultaneous} broadcast using dirty paper coding clearly outperforms static TDMA 
as shown in Figure \ref{fig:schemes3pairs}.

\item It is well known that the maximum multiplexing gain per pair of the half-duplex $K$ pair interference channel (with direct links) is $1/2$ \cite{Host-Madsen2005}. Interestingly, our results show that by adding a MIMO relay node in the network (and ignoring the direct link), one can significantly increase the multiplexing gain per pair in certain relevant scenario. For example, in the reciprocal channels scenario with transmit CSI at the relay, when $M\geq 2K$, each pair can achieve a maximum multiplexing gain of $1$. Even in the absence of transmit CSI, when $K=2$ and $M=2$, each one of the two pairs can achieve a maximum multiplexing gain of $2/3$ using the static DF-MAC-TDMA scheme. To the best of our knowledge, this is the first example of a multi-user network where adding a relay results in a larger multiplexing gain (i.e., pre-log factor).

\item While our analysis has focused solely on uni-cast traffic, one can generalize our results to the multi-cast scenario. In this case, the relay node will play the {\bf true} role of MIMO wireless switch. Currently, this generalized model is under our investigation.

\item One of the {\em subtle} advantages that the relay node offers in our set-up is a significantly reduced dependency on the available CSI. To illustrate this fact, let's compare it with the recently proposed interference alignment approach for the $K$ pair interference channel \cite{Cadambe2008}. This approach is the only known technique for achieving the optimal multiplexing gain per pair (i.e., $1/2$) in frequency/time selective interference channels. However, it requires {\bf global} knowledge about the network CSI at {\bf each node} in the network. In the MIMO switch setup, on the other hand, the nodes are only assumed to have {\bf local receive} CSI. Furthermore, for small networks, one can outperform the interference alignment scheme even when the relay node only has receive CSI and a relatively small number of antennas. In the case of large networks, the relay node needs a large number of transmit antennas and transmit CSI to achieve a multiplexing gain of $1$ per pair (this CSI requirement is still lower than the global CSI needed by interference alignment).

\end{enumerate}
\bibliographystyle{unsrt}
\bibliography{ref6}

\begin{thebibliography}{1}

\bibitem{Sendonaris2003a}
A.~Sendonaris, E.~Erkip, and B.~Aazhang.
\newblock User cooperation diversity. part i. system description.
\newblock {\em IEEE Trans. on Communications}, Vol. 51(11):pp. 1927--1938,
  November 2003.

\bibitem{Sendonaris2003}
A.~Sendonaris, E.~Erkip, and B.~Aazhang.
\newblock User cooperation diversity. part ii. implementation aspects and
  performance analysis.
\newblock {\em IEEE Trans. on Communications}, Vol. 51(11):pp. 1939--1948,
  November 2003.

\bibitem{Yuksel2007}
Melda Yuksel and Elza Erkip.
\newblock Multi-antenna cooperative wireless systems: A diversity-multiplexing
  tradeoff perspective.
\newblock {\em arXiv:cs/0609122v2}, July 2007.

\bibitem{Zheng2003}
Lizhong Zheng and D.~N.~C. Tse.
\newblock Diversity and multiplexing: a fundamental tradeoff in
  multiple-antenna channels.
\newblock {\em IEEE Trans. on Info. Theory}, Vol. 49(5):pp. 1073--1096, May
  2003.

\bibitem{Tse2004}
D.~N.~C. Tse, P.~Viswanath, and Lizhong Zheng.
\newblock Diversity-multiplexing tradeoff in multiple-access channels.
\newblock {\em IEEE Trans. on Info. Theory}, Vol. 50(9):pp. 1859--1874,
  September 2004.

\bibitem{JindalMay2004}
N.~Jindal, S.~Vishwanath, and Goldsmith A.
\newblock On the duality of gaussian multiple-access and broadcast channels.
\newblock {\em IEEE Trans. on Info. Theory}, Vol. 50(5):pp. 768--783, May 2004.

\bibitem{Caire2003}
G.~Caire and S.~Shamai~(Shitz).
\newblock On achievable rates in a multi-antenna broadcast downlink.
\newblock {\em IEEE Trans. on Info. Theory}, Vol. 49(No. 7):pp. 1691–--1706,
  July 2003.

\bibitem{Host-Madsen2005}
A.~Host-Madsen and A.~Nosratinia.
\newblock The multiplexing gain of wireless networks.
\newblock {\em International Symposium on Information Theory (ISIT)}, pages pp.
  2065--2069., Sept. 2005.

\bibitem{Cadambe2008}
Viveck~R. Cadambe and Syed~A. Jafar.
\newblock Interference alignment and the degrees of freedom for the k user
  interference channel.
\newblock {\em IEEE Trans. on Info. Theory}, Vol. 54:pp. 3425--3441, August
  2008.

\end{thebibliography}

\end{document}